# White Noise from the White Goods? Conceptual and Empirical Perspectives on Ambient Domestic Computing[1]

**Dr Lachlan D. Urquhart, LL. B, LL.M, Ph.D.**

Research Fellow in Information Technology Law, Horizon, Computer Science, University of Nottingham

Abstract:
Within this chapter we consider the emergence of ambient domestic computing systems, both conceptually and empirically. We critically assess visions of post-desktop computing, paying particular attention to one contemporary trend: the internet of things (IoT). We examine the contested nature of this term, looking at the historical trajectory of similar technologies, and the regulatory issues they can pose, particularly in the home. We also look to the emerging regulatory solution of privacy by design, unpacking practical challenges it faces. The novelty of our contribution stems from a turn to practice through a set of empirical perspectives. We present findings that document the practical experiences and viewpoints of leading experts in technology law and design.

Introduction

*"The house was full of dead bodies, it seemed. It felt like a mechanical cemetery. So silent. None of the humming hidden energy of machines waiting to function at the tap of a button."*
(Ray Bradbury, The Veldt, 1951)[2]

*"The house was an altar with ten thousand attendants, big, small, servicing, attending, in choirs. But the gods had gone away, and the ritual of the religion continued senselessly, uselessly."*
(Ray Bradbury, There Will Come Soft Rains, 1950)

Poetic portrayals forecasting the possible futures of home automation are not new. Ray Bradbury presciently demonstrates the darker dimensions in his two short works from the early 1950's, *The Veldt* and *There Will Come Soft Rains*[3]. The domestic Internet of Things (IoT) is the current favoured term, but draws on an extensive lineage of technological visions for the future of the home.

The longstanding utopian depiction of ambient domestic systems has been towards closer alignment of devices and services. The domestic IoT trend can be characterised as a networked ecosystem of intelligent products embedded in the social and physical infrastructure of the home. The devices are largely technically heterogeneous, each possessing different interfaces, sensing capabilities, networking protocols, and underlying goals. By utilising the sensing, monitoring and information sharing capabilities of different physical devices, distinct patterns of users' behaviour and daily life can be observed. Inferences can be drawn and used to provide contextually appropriate and adaptive value-added services, often within the mundane setting

---

[1] This chapter is based on the author's doctoral research completed at the University of Nottingham. The author is supported by the Horizon Centre for Doctoral Training at the University of Nottingham (RCUK Grant No. EP/G037574/1) and by the RCUK's Horizon Digital Economy Research Institute (RCUK Grant No. EP/G065802/1).

[2] As George Hadley continues to say to his wife… *"'...Lydia, it's off, and it stays off. And the whole damn house dies as of here and now. The more I see of the mess we've put ourselves in, the more it sickens me. We've been contemplating our mechanical, electronic navels for too long. My God, how we need a breath of honest air'"*
[3] Both in Bradbury, R. *Ray Bradbury Stories Volume 1* (New York: Harper Voyager, 2008)

of everyday practices. Importantly, the domestic setting is also heterogeneous, as homes are complex social spaces. Nevertheless, instead of a data driven cacophony of distinct artefact chatter, the goal is a harmonised end user experience.

One typical example might be a smart thermostat controlling room temperature by longitudinally observing patterns of occupancy. Another could be a smart fridge intelligently monitoring food stock to prevent wastage or suggest recipes. The interconnection of an array of devices through smart home ecosystems and platforms seeks to provide convenience and to optimise routine tasks for users. Such automation can involve input from the user, remotely controlling settings via mobile device applications, or increasingly, routine artificial intelligence enabled by machine learning capabilities. Further into the future, as automation increases, shifting interactions between users and systems can emerge, perhaps marked by software agents' performing tasks on a users' behalf, such scheduling cleaning cycles for washing machines.[4]

This framing of the march to the future has prompted much concern, especially as such systems are embedded within the intimate social context of the home. The link between the user needs and device functionality can be tenuous. From a regulatory perspective, countless recent news stories exemplify tricky issues emerging from across the privacy, information security and product safety law spectrum.

For privacy, smart TVs[5] and Barbie dolls[6] listening to conversations of home occupants have prompted discussions around privacy harms and adequate control over children's personal data. With security, there are search engines dedicated to finding unsecured internet connected baby monitors[7] and connected kettles leaking not water but WiFi passwords.[8] With physical safety concerns we see connected smoke alarms switching off when waved at[9] or learning thermostats randomly turning off heating[10]

Whilst these examples are illustrative, we want to systematically assess how ambient domestic systems and their regulatory challenges manifest in practice. Therefore, within this chapter, we couple our examination of conceptual literature with a turn to the practical experiences of experts from technology law and design. We look at both communities, as the solution to many of these regulatory issues is often a turn to the designers of technology. We aptly see this encapsulated in the notion of privacy by design (PbD).

Through our overtly multidisciplinary standpoint, we attempt to bring together computer science, particularly human computer interaction (HCI), and IT law. Before turning to our analysis, we now briefly introduce the empirical dimensions of our paper.

---

[4] Constanza, E. et al 'Doing Laundry with the Agents: A Field Trial of a Future Smart Energy System in the Home', (2014), *ACM SIGCHI '14,* 813-822
[5] BBC, 'Not in front of the Telly: Warning over 'listening' TV', (2015), *BBC Tech*, 9 Feb 2015, http://www.bbc.co.uk/news/technology-31296188
[6] Sheffield, J. 'Mattel's WiFi Barbie could be used to spy on children' (2015) *The Independent*, 18 March 2015. http://www.independent.co.uk/news/business/news/mattels-wifi-barbie-could-be-used-to-spy-on-children-10115724.html
[7] Porup, JM 'How to search the Internet of Things for photos of sleeping babies', (2016) *ARS Technica UK*, 19 Jan 2016. http://arstechnica.co.uk/security/2016/01/how-to-search-the-internet-of-things-for-photos-of-sleeping-babies/
[8] Pauli, D 'Connected kettles boil over, spill Wi-Fi passwords over London' (2015) *The Register*, 19 Oct 2015 http://www.theregister.co.uk/2015/10/19/bods_brew_ikettle_20_hack_plot_vulnerable_london_pots/
[9] Gibbs, S 'Bug in Nest Thermostat turns off heating for some' (2016), *The Guardian*, 15 Jan 2016 https://www.theguardian.com/technology/2016/jan/15/bug-nest-thermostat-turns-heating-off-for-some
[10] Arthur, C 'Nest halts sales of Protect smoke and carbon monoxide alarm on safety fears' (2014), *The Guardian*, 4 April 2014 https://www.theguardian.com/technology/2014/apr/04/nest-halts-sales-of-protect-smoke-and-carbon-monoxide-alarm-on-safety-fears

Our qualitative findings are based on thirteen detailed semi structured interviews conducted in Spring 2016[11] with leading UK experts in information technology law and design. We focus on experiences of thought leaders because their breadth of expertise provides us detailed insight into complex practical and strategic issues. We use pseudonyms to protect their identity and provide their years of experience, current position and areas of expertise instead (see Tables 1 and 2 below). We broadly cluster the participants under the labels of technologists and lawyers. In this chapter, we present how both communities navigate definitions and regulatory challenges of IoT, including the role of PbD as a solution.

Our six 'lawyers' have an average of 14 years of professional experience. They have a broad range of expertise across technology law including areas of: contracts, data protection (DP), intellectual property, software, e-commerce, accessibility, procurement, outsourcing, dispute resolution, and litigation.

Table 1.1: Legal Experts

| Pseudonym | Job | Years of Experience | Specialism |
|---|---|---|---|
| Blair | Managing Director and Lawyer | 9 years | IT and Telecoms Law |
| Campbell | Full Time Academic | 8 years | Teacher & Researcher in Law |
| Duncan | Partner | 25 Years | Technology and public procurement law |
| Ewan | Partner | 14 years | Technology, Data Protection and Information law |
| Findlay | Consultant | 20 years | Privacy and Information management policy |
| Innes | Legal Associate | 8 years | Commercial Technology, Intellectual Property, and Data Protection Law |

The seven 'technologists' have an average of 32 years of professional experience at both strategic and operational levels. Their expertise draws on specialisms like wireless networking, information security, privacy and identity, data science, ethics, big data, telecoms, cloud computing, interaction design, and digital media.

Table 1.2: Technology Experts

| Pseudonym | Job | Years of Experience | Specialism |
|---|---|---|---|
| Allan | Director & Technical Professional | 32 years | Digital Identity & Privacy |
| Iain | Professor & Chair | 25 years | Interaction Design and Digital Media |
| Jess | Senior Level Researcher | 28 years | Cybersecurity, big data and ethics of data science |
| Gordon | Vice President & Primary Consultant | 30 years | Wireless networking and sensors |
| Kenneth | Vice President & Visiting Professor | 42 years (telecoms)/15 years (visiting professor) | Engineering and Telecoms |

---

[11] The average length of interview was **44 minutes**. The participants were given an information sheet and consent form to sign, prior to the interview, and this study passed through the University of Nottingham Computer Science ethics approval process. The interviews were audio recorded, transcribed verbatim, coded and analysed using **thematic analysis following** Braun, V. and Clarke, V. (2006) Using thematic analysis in psychology *Qualitative Research in Psychology*, 3 (2). pp. 77-101.

| Magnus | Chief Technology Officer | 40 years | Wireless technologies and smart devices |
| Hamish | Managing Consultant | 30 years | Cybersecurity & identity management |

In terms of structure, in Part I we focus on the nature of the domestic internet of things, starting with the history of ambient domestic computing, issues around such technological visions, before looking to current conceptual and empirical perspectives on the IoT. In Part II, we look at the regulatory challenges surrounding such systems, again in theory and practice. We focus in particular on privacy and data protection concerns, particularly around managing flows of personal information and obtaining consent. In Part III we explore the regulatory solution of privacy by design, considering the legal basis, concerns about the concept and how it currently manifests in practice. We conclude by with reflections on the contemporary nature of PbD for the IoT.

## Part I: The Nature of Domestic IoT

### a) The Development of Ambient Domestic Computing

We now look at how ambient domestic computing has emerged over the past 25 years. Weiser's archetypal vision of ubiquitous computing is a key milestone in the research agenda of post-desktop computing.[12] With ubicomp, systems have disappeared and *"weave themselves into the fabric of everyday life until they are indistinguishable from it"*.[13] Satyanarayanan's later vision of 'pervasive computing' also considers invisibility in use, where *"a pervasive computing environment as one saturated with computing and communication capability, yet so gracefully integrated with users that it becomes a ''technology that disappears"*.[14] Such ubiquity requires computing to be managed appropriately, to become 'calm' because *"if computers are everywhere they better stay out of the way, and that means designing them so that the people being shared by the computers remain serene and in control'"*.[15] Implementing this vision relies on building in contextual awareness, indeed as Dourish states, *"when computation is moved 'off the desktop' then we suddenly need to keep track of where it has gone"*.[16] However, as Rogers argues, achieving 'context aware computing' requires engineering approaches that enable *"detecting, identifying and locating people's movements, routines or*

---

[12] Weiser, M 'The Computer for the 21st Century' (1991) *Scientific American*, 94-104;

Weiser, M & Brown, JS 'The Coming Age of Calm Technology' In Denning, PJ and Metcalfe, R.M *Beyond Calculation,* (New York: Copernicus, 1997) p1-2;

Grudin, J 'The Computer Reaches out: The Historical Continuity of Interface Design' *Proceedings SIGCHI Conference Human Factors in Computer Systems (CHI'90)* (New York: ACM Press, 1990) 261-268.

*With* Grudin (1990), *Ubicomp*, that is, many computers to one person was framed as the next trend in the transition from *mainframe computers* (many people to one computer) to *personal computers* (one person to one computer), via the transition of the *internet/distributed computing.*

[13] Weiser, 1991, p94
[14] Satyanarayanan, M 'Pervasive Computing: Visions and Challenges' (2001) *IEEE Personal Communications* 8(4) p2
[15] Weiser & Seely Brown, 1997, p78
[16] Dourish, P 'What We Talk About When We Talk About Context' (2004) *Personal and Ubiquitous Computing* 8(1) 19-30, p20

*actions with a view to using this information to provide relevant information that may augment or assist a person or persons".*[17]

Significant industry and government investment attempted to bring these visions mainstream, as typified in the European Commission's 'Disappearing Computer programme'[18] and Philip's 'Vision of the Future'.[19] A major crossover stream of work was Ambient Intelligence (AmI) a vision defined by five features: embedded, context aware, personalised, adaptive and anticipative systems.[20] Indeed, as Lindwer et al argue, AmI is "*the vision that technology will become invisible, embedded in our natural surroundings, present whenever we need it, enabled by simple and effortless interactions, attuned to all our senses, adaptive to users and context and autonomously acting. High quality information and content must be available to any user, anywhere, at any time, and on any device*".[21]

Such utopian forecasts have been criticised from a number of perspectives over the years. Reeves (2012) argues future orientated, quasi fictional technological visions, whilst dominant in computing, are not predictions but merely a commentary of the present.[22] They often never materialise, as some argue is the case for Weiser's 'ubicomp' after 25 years.[23] Contrastingly, Bell and Dourish argue ubicomp is here, just not the 'yesterday's tomorrow' Weiser envisaged.[24] As opposed to his clean, seamlessly networked[25] technological future, an alternate present has appeared, one that is seen not in labs, but in the real world.

The harm of committing to future visions is the present, and the difficult challenges therein, can be ignored. As Bell and Dourish argue, in the vision of engineering seamless networking (that is, no gaps in coverage), the "*…messy present can be ignored, although infrastructure is always unevenly distributed, always messy. An indefinitely postponed Ubicomp future is one that need never take account of this complexity*".[26] Furthermore, change takes time, and as HCI has long recognised, the smart home will not emerge overnight and as Edwards and Grinter state "*new technologies will be brought piecemeal into the home*".[27]

Accordingly, instead of engineering a grand vision, there is a shift away from focusing on how to implement canonical underpinning principles, like calmness[28] or invisibility.[29] Instead the user, and how technologies manifest in practice, is key. A good example is Weiser's invisibility. As Tolmie et al argue, this is not just physical invisibility, but instead computing

---

[17] Rogers Y. 'Moving on from Weiser's Vision of Calm Computing: Engaging Ubicomp Experiences' in *Proceedings of the 8th International Conference on Ubiquitous Computing (UbiComp'06)* (New York: ACM Press, 2006) 404-421, p408
[18] http://www.disappearing-computer.eu/resources.html
[19] http://www.research.philips.com/technologies/download/homelab_365.pdf
[20] Aarts, E & Marzano, S *The New Everyday: Views on Ambient Intelligence* (Rotterdam: 010 Publishers 2003)
[21] Lindwer, M, et al 'Ambient Intelligence visions and achievements: Linking abstract ideas to real-world concepts', (2003) *Design, Automation and Test in Europe Conference*, p1
[22] Reeves, S 'Envisioning Ubiquitous Computing' *In Proceedings SIGCHI Conference Human Factors in Computer Systems (CHI'12)* (New York: ACM Press 2012) 1573 -1582, p1580
[23] Caceres, R & Friday A 'Ubicomp Systems at 20: Progress, Opportunities and Challenges' (2012) In *IEEE Pervasive Computing* 11(1), 14-21, p15
[24] Bell, G., & Dourish, P. 'Yesterday's Tomorrow's: Notes on Ubiquitous Computing's Dominant Vision' (2006) *Personal and Ubiquitous Computing* 11(2), 133-143.
[25] Where there are no issues with connectivity or networking for devices, seamless interactions for users with systems
[26] Bell and Dourish, 2006, p140
[27] Edwards, K & Grinter R 'At Home with Ubiquitous Computing: Seven Challenges' *Proceedings of the 3rd International Conference on Ubiquitous Computing (UbiComp'01)* (New York: ACM Press, 2001) 256-272, p257
[28] Rogers, 2006, p406
[29] Tolmie, P. et al 'Unremarkable Computing' *In Proceedings SIGCHI Conference Human Factors in Computer Systems (CHI'12)* (New York: ACM Press, 2012)

needs to become so routine in life it is no longer 'remarkable'. Creating such unremarkable systems requires situated understanding of the social context of use, the actions and routines of daily life, including what makes activities routine.[30]

Similarly, technically driven smart home research has long contented benefits to users of increased efficiency, comfort, convenience, energy management, care, security.[31] As Wilson et al argue, here too designers need to look at the user, to see "*how the use and meaning of technologies will be socially constructed and iteratively negotiated, rather than being the inevitable outcome of assumed functional benefits*".[32] This requires recognising homes are '*internally differentiated, emotionally loaded, shared and contested places'*[33], and as Leppänen and Jokinen state "inhabitants themselves make a home and little everyday practices make the known life go on...a smart home should not be smarter than its inhabitants".[34]

Empirical evidence on the impact of smart thermostats, CCTV or locks unpacks how smart home technologies can mediate end users lives. Ur et al, for example, found parental auditing of home entry/exit through smart locks and cameras, whilst convenient and safer, impacted trust relationships with their children.[35] Domestic sensing often leads to complex trade-offs between observers knowing observed family members are safe and protected against the observed members' perceptions of spying.[36] Occupants can become accustomed to monitoring technologies and adjust their behaviour accordingly.[37] Technologies can become 'unremarkable' over time too. For example, provided smart thermostats work properly, they become mundane over time, and earlier home occupant frustration, lack of comprehension and concerns of control over functionality fade away.[38]

Users, their social context, needs, relationships and environment need to be positioned at the core of design. Prescriptive engineering principles within near future technological visions can cause these to be neglected. From a regulatory perspective, looking to users and how technologies impact their interests is important, but we need to look at current, as opposed to future visions. Accordingly, we now explore what IoT is descriptively, by turning to current

---

[30] Tolmie et al, 2002, p402 "*An orderly aspect of things with a routine character is that they can serve as resources for the mutual coordination of unremarkable activities…these resources are mutually available and mutually accountable for those involved in the routine. Also things do of course go wrong in domestic life, alarms can fail – but failure, in contrast to accomplishment, is remarkable and the elements held to account when part of a routine fails are the very ones that are unremarkable at other times*"
[31] Wilson, C. et al 'Smart Homes and Their Users: A Systematic Analysis and Key Challenges' (2015) *Personal and Ubiquitous Computing* 19. 463-476
[32] Wilson et al, 2015, p466
[33] Wilson et al, 2015, p470
[34] Leppänen, S & Jokinen, M. 'Daily Routines and Means of Communication in a Smart Home', in R Harper *Inside the Smart Home*, (London: Springer Verlag, 2003) p223
[35] Ur, B, Jung, J & Schechter, S 'Intruders Versus Intrusiveness: Teens' and Parents' Perspectives on Home- Entryway Surveillance' *Proceedings of the international conference on Ubiquitous Computing (UbiComp'14)* (New York: ACM Press, 2014)
[36] Mäkinen, L 'Surveillance On/Off: Examining Home Surveillance Systems From The User's Perspective', (2016) *Surveillance & Society* 14(1): 59-77;
Choe, E.K. et al, 'Living in a Glass House: A Survey of Private Moments in the Home', *Proceedings of the international conference on Ubiquitous Computing (UbiComp'11)* (New York: ACM Press, 2011)
[37] Oulasvirta, A. et al 'Long-term Effects of Ubiquitous Surveillance in the Home', *Proceedings of the International Conference on Ubiquitous Computing (UbiComp'12)* (New York: ACM Press, 2012)
[38] Yang, R. and Newman, M (2013) Learning from a Learning Thermostat: Lessons for Intelligent Systems for the Home, *Proceedings of the International Conference on Ubiquitous Computing (UbiComp'13)* (New York: ACM Press, 2013)
Yang et al (2014) Making Sustainability Sustainable: Challenges in the Design of Eco-Interaction Technologies, *Proceedings SIGCHI Conference Human Factors in Computer Systems (CHI'14)* (New York: ACM Press, 2014) perspectives on the term. We seek to cut through the fiction in order to navigate the practical regulatory challenges it poses, and to situate how domestic IoT is framed in practice currently.

b) Exploring the Emergence of domestic IoT: Conceptual and Empirical Perspectives

Divining clarity around IoT is tricky as we find it sitting at the summit of the 'peak of inflated expectations', clouded in hype and optimism.[39] Famously, Cisco predict 24 billon internet connected devices by 2019[40] and Huawei 100 billion by 2025.[41] Similarly, the OECD foresee a family of four will own seven smart light bulbs, five internet connected power sockets, one intelligent thermostat and so on by 2022.[42]

Unlike AmI, Ubicomp, or Pervasive Computing, IoT generally lacks the similar canonical technical framing.[43] When Ashton first used the term in 1999[44] he focused on tracking objects via machines instead of humans in a product supply chain. Since then, IoT has emerged in a broad range of application domains, from the built environment of smart homes and cities, to smart energy grids, intelligent mobility through connected and autonomous vehicles, and smart healthcare through wearables and the quantified self. [45]

To get a handle on how the term is popularly understood, we look to perspectives from different stakeholders. We find that by considering the UK Government Office for Science[46]; EU Article 29 Working Party[47]; Cisco[48]; UN International Telecoms Union[49], Internet Engineering Task

---

[39] Gartner Hype Cycle for Emerging Technologies (2015) https://www.gartner.com/newsroom/id/3114217
[40] Cisco Website, *Visual Networking Index,* (2016) available at http://www.cisco.com/c/en/us/solutions/service-provider/visual-networking-index-vni/index.html
[41] Huawei Website, *Global Connectivity Index*, (2016)
[42] Working Party on Communication Infrastructures and Services Policy, *Building Blocks for Smart Networks*, (Paris: OECD, 2013)
https://www.oecd.org/officialdocuments/publicdisplaydocumentpdf/?cote=DSTI/ICCP/CISP(2012)3/FINAL&docLanguage=En
[43] McAuley, D. 'What is IoT? That is not the Question', (2016) *IoT UK*, Accessed at http://iotuk.org.uk/what-is-iot-that-is-not-the-question/ advises
against focusing on the lack of fundamental technical definition for IoT because "*IoT is not about technical capabilities or novelty, rather it is a social phenomenon that reflects a significant proportion of society, and importantly businesses, who have started to recognise that there is value in building a virtual presence for many of our everyday physical things*" (p1).
[44] Ashton, K. (2009), That Internet of Things Thing, *RFID Journal* at http://www.itrco.jp/libraries/RFIDjournal-That%20Internet%20of%20Things%20Thing.pdf
[45] Walport M, *Internet of Things: Making the Most of the Second Digital Revolution*, (London: UK Government Office of Science 2014) p9-11
[46] Walport, 2014, p13 - IoT "is made up of hardware and software technologies. The hardware consists of the connected devices – which range from simple sensors to smartphones and wearable devices – and the networks that link them, such as 4G Long-Term Evolution, Wi-Fi and Bluetooth. Software components include data storage platforms and analytics programmes that present information to users"
[47] Article 29 Working Party, *Opinion 8/2014 on the Recent Developments on the Internet of Things WP 23*, (Brussels: European Commission, 2014) – IoT is devices "that can be controlled remotely over the internet…most home automation devices are constantly connected and may transmit data back to the manufacturer" see s1.3
[48] Cisco, *The Internet of Everything* (San Jose: Cisco, 2015) at http://www.cisco.com/c/dam/en_us/about/business-insights/docs/ioe-value-at-stake-public-sector-analysis-faq.pdf the IoE is a "*networked connection of people, process, data, and things*" p1
[49] International Telecommunications Union, *Overview of the Internet of Things,* (Geneva: ITU, 2012) IoT is "*a global infrastructure for the information society, enabling advanced services by interconnecting (physical and virtual) things based on existing and evolving interoperable information and communication,*" where a 'thing' is: an "*object of the physical world (physical things) or the information world (virtual things), which is capable of being identified and integrated into communication networks*" p1

Force[50] and cross disciplinary academic group, Cambridge Public Policy[51] there are a wide range of descriptive attributes assigned to IoT, including:

- Physical objects with a digital presence,
- Socially embedded,
- Remotely controllable,
- Constantly connected with networking for information sharing between people, processes and objects,
- Surrounded by an ecosystem of stakeholders interested in the personal data supply chain for example, third parties,
- Tied to backend computational infrastructure (for example, cloud, databases, servers),
- Device to device/backend communication often without direct human input

We want to see to what extent these attributes also play out in our empirical findings. In turning to our lawyers, we also see IoT being framed quite openly. Beyond traditional notions of computing, IoT is viewed as any networked sensor or device. As Campbell [University Reader, 9 years, Media and Tech Law] contends, IoT is *"...things that aren't computers that are connected to the internet. Computers in the sense of desktop, laptop, higher spec mobile devices, basically everything that is not one of them but is somehow internet connected...."*. Notwithstanding, more mundane, established systems, like the smart phone, are still considered by some to be IoT as they mediate end user interactions with other IoT devices and services. As Findlay [Consultant, 20 years, Privacy and Information Management] maintains: *"more and more digital services and devices are being created where the smart phone is the hub if you like, and the router in your home is the hub to your connected life."*

The legal community tend to contextualise IoT using illustrative examples from a number of application areas, often drawn from their own practical experience from their roles as advisers. These include:

- machine to machine (M2M) industrial and retail applications,
- consumer products like autonomous cars,
- wearables like fitness bands,
- a range of smart home systems for security, lighting, energy management, entertainment and comfort.

Overall, when turning to the technologists we see tighter but more contested definitions of IoT. Like the lawyers, they also classify IoT by reference to different applications and sectors, with some arguing IoT is just the next hyped technology trend, like Web 2.0, Cloud Computing, M2M or connected devices that went before. Magnus [Chief Technology Officer, 40 years, Wireless Technology and Smart Devices] captures this well:

*"they changed name from connected devices to M2M to internet of things – the cynical would say when something doesn't take off you just change the name –and it hasn't really taken off,*

---

[50] Arkko, J et al, *IETF RC 7452: Architectural Considerations in Smart Object Networking,* (Fremont, Internet Engineering Task Force, 2015)
IoT is a "trend where a large number of embedded devices employ communications services offered by the Internet protocols. Many of these devices often called smart objects are <u>not directly operated by humans</u>, but exist as components in buildings or vehicles, or are spread out in the environment" p1

[51] Deakin, S. et al, *The Internet of Things: Shaping Our Future*, (Cambridge: Cambridge Public Policy, 2015)
 "<u>sensors</u> which react to physical signals; software in these sensors <u>transmitting information</u>; a <u>network</u> for this information to be transmitted on; a database and control system which <u>receives and processes</u> this data, and <u>sends a message back</u> out over the network to instruct the initial device or another one that is networked" p8 *the IoT is still predominantly hype… over the years [I've] ended up helping people putting wireless into everything from sex toys to snow ploughs*." [Magnus, CTO]

Broadly, the technologists focus on more technical attributes, such as the physicality of objects, embedded computation, pervasiveness of communications infrastructure, global nature of internet connectivity and variations in user interface. Allan's [Director, 28 years, Digital Identity & Online Privacy] definition captures these various elements well, positing IoT is:

"*a device that has processing capability that has some degree of user interface but where the user interface is partial,* in other words, … if you have a smart lightbulb, part of the user interface would be you can turn it on and off, part of it might be you can remotely control times that it goes on and off, but that is not all the functionality the device has, so it also might have network capability that you don't have an interface to, it's almost certainly got backend cloud communication capabilities, and data transfer capabilities that you're very unlikely to have an interface to." [Allan, Director]

However, for some, framing IoT around the nature of objects or networking is misguided. Instead, they want to go beyond just the vision of a 'connected product' to a more holistic vision. They focus on the user, their data flows and various practices around a technology. As Iain [Professor, 25 years, Interaction Design and Digital Art] puts it, IoT is:

"…about *flows of data through practices, and most social personal practices involve physical artefacts, environments and other people*, they are not just selves, data is often co-produced making it hard to identify ownership, it tends to be ubiquitous in a networking sense. Geography has become very complicated, it is very difficult to understand a geographic boundary around a thing, practice or dataset. The bug for me is to assume the things are objects, whereas things are far more complicated concept… [it's about] *an internet of practice, but you can't practice without some objects, without other people.*" [Iain, Professor]

Some technologists stress the centrality of data as a commodity, as opposed to the user, due to the importance of data for analysis and creation of new services. These participants were also wary of regulation, insofar as it may limit their access to data, as Magnus puts it:

"…IoT is about taking data off a device and then doing data analytics on that somewhere in the cloud. The stuff in the middle doesn't really matter it just needs to be there and work…from my mind where I'm involved in a lot of data analytics on large databases, people always say, 'oh god data overload is terrible', I think it is an aspiration, I want more data, the more I have, the more I can do with it. *Somebody tells me to minimise my data is basically trying to restrict a business model.*" [Magnus, CTO]

Overall, in contrast to ubicomp, or AmI, we see a less canonical narrative emerging around IoT. Whilst hype persists, there is no single, unifying vision of what is or is not IoT. Conceptually and empirically, a more flexible framing is emerging. The strong focus on IoT applications sidesteps the need to fixate over where the margins of IoT lie. Most importantly, with applications, like the home, come contexts of use and end users who have various needs and interests. By considering this level, as opposed to a grand vision, a richer, situated vision of IoT can emerge, looking at the practices, interactions and relationships end users have with the technologies.

Stemming from these findings, we now consider regulatory implications IoT can pose, beginning with general challenges before shifting to privacy orientated perspectives.

Part II: Regulatory Challenges of IoT

a) General Regulatory Challenges of IoT

Conceptually, we observe a range of IoT regulatory concerns discussed in the literature. These include:
- Lack of interoperability between devices and across platforms[52]
- Market dominance and inadequate competition around firms[53]
- Insufficient spectrum and internet protocol (IP) addresses for devices (IPv6 solves much of this)[54]
- Lack of leadership on industry standards[55]
- Responsibility and liability for harm[56]
- Technical education, appropriate regulation and trust in the security of these systems.[57]

In practice too, both the lawyers and technologists provide a picture of overarching regulatory issues they have experienced. These primarily include safety, liability and responsibility for harm, data security, intellectual property, funding and interoperability.

More generally, the technologists question the fitness for purpose of regulatory frameworks and legislation, particularly for how IoT impacts existing legal principles and consumer rights. As Gordon [Principal Consultant, 30 Years, Wireless Networking and Sensors] frames it, "*fundamentally it is such a new tech, new area, it is fairly wild west*" and in some contexts, this is more apparent than others. A good example is the contrast between safety and security. Whilst devices strict market access controls around electrical safety is enforced, similar oversight has not emerged for IoT security. Interestingly, security, as opposed to privacy, often emerges as a more legitimate concern for the technologists. Some feel data protection is limiting business models, stifling innovation or creative practice, and instead regulating misuse of personal data should be favoured. For them, good security practices may require focusing on the diligence of designers to use best practices as opposed to post hoc responses like insurance for security breaches. However, this requires systematic consideration of IoT security risks in design, and this needs to be framed current state of the IoT industry.

In particular, the embryonic market, the heterogeneity of the device ecosystem, and lack of industry standards complicate any application of regulations. Again, for Gordon, "*the [IoT] industry itself is very immature, very young. I've described it before as a 'primordial soup' there are a few things that have crawled out on land, some are slithering about, some have got legs, some have arms but we do not have many fully formed creatures yet*" [Principal Consultant]. Accordingly, dominant industry standards are yet to crystallise. The experts contend the commercial process of establishing dominant platforms or communications protocols is in progress. However, the technical heterogeneity of the IoT ecosystem means the standards landscape is likely to remain unsettled for some time. Long term, the technologists argue that lack of device interoperability may enable market dominance by actors with the resources to invest in setting commercial standards and minimising competition from smaller

---

[52] Deakin et al, 2014, p7
[53] Brown, I, *GSR Discussion Paper: Regulation and the Internet of Things,* (Geneva: International Telecommunications Union, 2015), p19
[54] Brown, 2015, p19
[55] Bouverot, A *GSMA: The Impact of the Internet of Things - The Connected Home,* (Barcelona: GSMA, 2015) at http://www.gsma.com/newsroom/wp-content/uploads/15625-Connected-Living-Report.pdf
[56] Rose, K. et al, *Internet of Things: An Overview*, (Geneva: Internet Society, 2015) p38
[57] Walport, 2014

actors. Similarly, whilst communications protocols like Thread, Weave, Z-Wave ZigBee are emerging, legally there remains device interoperability concerns, as Duncan [Partner, 25 years, Technology and Procurement Law] contends:

"[legally] …*it's always quite difficult in terms of the standardisation process, for it [IoT] to work together there have got to be effective standards to make everything work and at the moment, obviously everybody is jostling to become the hub, so that they want their thing to be in the driving seat, to be the heart of what is going on in the IoT, and obviously time will tell who will be the winners and who will be the losers.*" [Duncan, Partner]

The heterogeneity of the IoT ecosystem impacts device users too. The technologists are concerned about how variations in device interfaces shape the end user interactions. Allan argues, "*there is a general direction of travel you can plot a line from browsers to smart phones to devices that run an app to ambient devices. At each stage along that line, the end user gets access to a more restricted user interface, and gets fewer and fewer controls over what the device is actually doing, and in the case of ambient, passive collection is at zero interface*" [Allan, Director]. Furthermore, devices mediate actions of end users and variations in customisability of devices can shape control and choice in a subtle, everyday ways. As Blair [Managing Director of Law Firm, 9 years, IT and Telecoms Law] puts it "…*at what point do you [as a consumer] find that actually you are not benefitting from all this wonderful technology but you're actually living in an environment where it is your fridge telling you when you can or cannot eat, because you've exceeded the number of times you've opened and closed the door in a day…*" [Blair, Managing Director]

To conclude, we see from these experiences that the experts engage with a wide range of regulatory issues around IoT, where safety and security are particular concerns. Furthermore, the embryonic nature of the IoT, the heterogeneity of the device ecosystem, and lack of industry standards further complicate the regulatory landscape. We now look at both conceptual and empirical perspectives, particularly prominent concerns around privacy and DP.

b) Data Protection and Privacy Challenges of IoT

Predecessors to the domestic IoT have long prompted reflection on privacy challenges for end users.[58] With Ubicomp, Čas (2009) worries "*ubiquitous computing will erode all central pillars of current privacy protection*" (p167) and reconciling ubicomp benefits with privacy risks is a considerable challenge. Spiekermann and Pallas fear paternalism through ubicomp, where non-negotiable binary rules enable automatic compliance, limit control and reduce user autonomy.[59] With AmI, the SWAMI[60] project Wright et al systematically consider a multitude of threats[61] and vulnerabilities[62] highlighting privacy, security and trust issues from technical, regulatory and socio-economic perspectives.[63] With IoT, such concerns persist[64] and we focus here on two primary areas of regulatory concern for IoT, namely: managing flows of personal information and user consent.

---

[58] Belotti, V.,& Sellen, A 'Design for Privacy in Ubiquitous Computing Environments' (1993) *ECSCW '93*, 77-92
Čas, J. 'Ubiquitous Computing, Privacy and Data Protection' in Gutwirth, S. et al *Computers, Privacy and Data Protection: An Element of Choice,* (Netherlands: Springer, 2009)
Wright, D. et al *Safeguards in a World of Ambient Intelligence*, (Netherlands: Springer, 2008)
De Hert, P. et al 'Legal Safeguards for Privacy and Data Protection in Ambient Intelligence', (2009) *Personal and Ubiquitous Computing* 435-444
[59] Spiekermann, S. and Pallas, F. 'Wider Implications of Ubiquitous Computing', (2005) *Poiesis & Praxis: International Journal of Ethics of Science and Technology Assessment* 4(1), 6-18
[60] Safeguards in a World of Ambient Intelligence
[61] Chapter 4.6 – eg 'lack of transparency'; 'loss of control and technology paternalism'; 'system complexity, false positives and unpredictable failures'
[62] Chapter 4.7

### i) Managing Flows of Personal Information

IoT ecosystems involve flows of information between different devices, users and services. The setting of the home is key. Brown (2015) argues IoT as problematic because it exists in private domestic contexts, presenting an attack target that is harder to secure and can compromise physical safety.[65] Indeed, as Rosner states, "*it is not the Internet of Things that raises hackles – it is the Intimacy of Things*".[66] Profiling is another big concern for IoT.[67] The A29 WP worries detailed inferences can be drawn about daily life where "*analysis of usage patterns in such a context is likely to reveal the inhabitants' lifestyle details, habits or choices or simply their presence at home*".[68] Similarly, Deakin et al. note combinations of non-personal data may create sensitive personal data (which consequently need explicit user consent), for example, systems that collect "*data on food purchases (fridge to supermarket system) of an individual combined with the times of day they leave the house (house sensors to alarm system) might reveal their religion*"[69]

IoT concerns needs to be situated against the wider European climate of user unease around control of their personal data. In a 2015 Eurobarometer Survey of c.28,000 EU citizens' attitudes to personal DP, two-thirds of respondents are "*concerned about not having complete control over the information they provide online*".[70] Nearly 70% think both prior explicit approval is necessary before data collection and processing, and worry about data being used for purposes different from those at collection.[71] Around 60% distrust telecoms firms, internet service providers and online businesses.[72] Looking to IoT more specifically, a recent global study by 25 DP regulators of IoT devices shows "*59 per cent of devices failed to adequately explain to customers how their personal information was collected, used and disclosed… [and] … 72 per cent failed to explain how customers could delete their information off the device*"[73]

Against this backdrop, we now turn to our empirical findings. The complexity and diversity of device, service and user interactions can make it hard to comprehend the flows of information, and the rationales behind them. As Iain puts it "*Across the IoT then we will have to deal with a whole host of transactions, and some of those will be incredibly small, and involve small forms of currency, not just economic, but data, if a kettle talks to the fridge, or the toothbrush*

---

[63] p269 "*This book has identified many threats and vulnerabilities and many safeguards for dealing with them. Perhaps we have identified too many safeguards or made too many recommendations, at least, in the sense that so many may seem daunting*"
[64] Edwards, L. 'Privacy, Security and Data Protection in Smart Cities: A Critical EU Law Perspective' (2016) *European Data Protection Law Review*, 2(1):28-58; Rosner, G. *Privacy and The Internet of Things* (Sebastopol: O'Reilly, 2016); Peppet, S. 'Regulating the Internet of Things: First Steps Toward Managing Discrimination, Privacy, Security, and Consent' (2014) *Texas Law Review* 93(1):87-176.; Weber, R. 'Internet of Things- New Security and Privacy Challenges', (2010) *Computer Law and Security Review*, 26(1), 23-30
[65] Brown, 2015, p25
[66] Rosner, 2016, p18
[67] Article 29 Working Party, 2014, p8
[68] Article 29 Working Party, 2014, p6-8
[69] Deakin et al, 2015, p15
[70] European Commission, *Special Eurobarometer 431 'Data Protection'* (2015) available at http://ec.europa.eu/justice/newsroom/data-protection/news/240615_en.htm p6
[71] European Commission, 2015, p58
[72] European Commission, 2015, p63
[73] ICO, 'Privacy regulators study finds Internet of Things shortfalls' (2016) *ICO Blog*, 22 Sept 2016 https://ico.org.uk/about-the-ico/news-and-events/news-and-blogs/2016/09/privacy-regulators-study-finds-internet-of-things-shortfalls/

*to the toothpaste, or the door to the car, will many of those things...how will they want to be private, how will they construct valuable experiences? [it] will be really really challenging.*" [Iain, Professor]

Nevertheless, better understanding of the personal data flows within these micro transactions needs to emerge. We found one practical challenge is understanding who is legally responsible for the information and mapping their obligations therein. This requires asking practical questions like what is data being used for, by whom, where it is being stored and how long it is being kept, but as Innes argues, the breadth of stakeholders, platforms and applications make this a tricky exercise:

"*I think identifying who the data controller is, particularly when not just speaking about a single device but in an interconnected environment where you have all these providers of different tech hosted on different platforms maybe a building landlord who is responsible for a building then you have got tenants and service providers and it gets very complicated with all these different players who are maybe using this data for their own different purposes, you start seeing different levels of data controllers for different obligations. It's going to get very complicated*". [Innes, Associate]

A particular concern both technologists and lawyers flag is around flows of data to third parties. Whilst interactions between users and primary service providers may be apparent, and legitimate, protecting user rights around third parties is harder. As Allan argues, "*you have a right to know that a third party has data about you and that right is kind of implied by the right to see that data, and to correct it and so on, if you simply don't know or you don't know who the third party is or how to get in touch with them, then your ability to exercise that right is completely undermined*" [Allan, Director]

Indeed, end users have a range of DP rights, but the challenges in establishing who is responsible for them can impact how they are protected and realised in practice. One reason these rights are so important is the control they afford the end user over their personal data. Control in this context is not just about the data itself, but also controlling the inferences that can be drawn from the data, in particular any prejudicial impacts. Blair's example captures this well, depicting a hypothetical scenario where insurers use wearables to monitor user activity to vary insurance premium rates:

"*the idea of automated decision taking, and making, based on the data acquired from IoT... take for example, private health insurance and they say here you go, we will send you one of our smart pedometers or our smart fitness trackers, you wear it, as part of your contract of health insurance, and we will price your premium based on the level of activity that we see you doing, if we see you sitting in your chair all day you will get a higher premium because you're inactive, if we sense you play rugby you will get a higher premium because you will hurt yourself...*" [Blair, Managing Director]

To conclude, in practice establishing responsibility for flows of data is a key challenge, but often frustrated by the complexity of data flows in the IoT. We now turn to a mechanism that seeks to increase control over data for users: consent.

### ii) Consent

Whilst it is not the only legal grounds for processing personal data, consent is an important one, especially for sensitive personal data. For IoT, the users' insufficient knowledge of data processing by physical objects, inadequate consent mechanisms [74] and lack of control over data sharing between such objects are key concerns.[75] As Edwards has argued "…*even if methods can be found for giving some kind of notice/information, the consents obtained in the IoT are almost always going to be illusory or at best low-quality in terms of the EU legal demand for freely given, specific and informed consent.*"[76]

In our experts' experience, obtaining freely given, informed end user consent with IoT can be challenging, especially when users are unaware of the nature of data collection. As Allan argues, "*consent is being tampered with, it's being assumed in some cases because the default setting for many devices may be that they connect and communicate, whether they ask or not, and consent is also being undermined because you don't necessarily know what data it is collecting or sharing, and you don't know what is being done with the data.*" [Allan, Director]

The requirement to inform end users about data collection is impacted by device heterogeneity. The variations in IoT device interfaces necessitate more creative mechanisms for delivering information during the consent giving process. As discussed above, smart phones play a key intermediary role in the IoT ecosystem, and can be a conduit for information, provided it belongs to the end user. Findlay outlines an approach:

"*In the IoT the challenge is many devices don't have a user interface, like the Nest smart thermostat. It is the smartphone, the web and email which you are using. You have two things right, your authentication device, the thermostat, but you can't present that info on the authentication device, so you need a consumption device like a smartphone, laptop or desktop, so it is going to require some thinking there, particularly when the law also says you also have to secure evidence of consent. So that needs to give some rise. So when it comes to an IoT device, and then the consumption device for information, how do you know if the consumption device is mine.*" [Findlay, Consultant]

Another attribute of consent is it is meant to be freely given. However, when terms of service change and renewed user consent is required, negotiation is lacking. Consumers are faced with a choice of either accepting changes or to stop using the product, and power asymmetries between consumers and IoT product/service providers quickly become evident. Blair questions this practice, stating "*the idea that something is freely given, when you've paid £250 on your smart thermostat, the idea of use it or lose it because of a change of terms makes it very questionable if any consent is ever freely given.*" [Blair, Managing Director]

A connected issue is how changes in consent manifest across different devices in an IoT ecosystem. The practicalities of designing cross device consent making processes requires reflection and tailoring to different contexts and end users. One consideration is that IoT devices often operate in settings where consent of multiple end users is required. Lawyers and designers need to create approaches that provide notification of data processing and obtain consent from all data subjects affected, for example visitors to the home being captured by a domestic security system. Innes [Associate, 8 years, Commercial IT and Data Protection Law] frames this challenge by contrasting consent mechanisms for a personal fitness band with one user and a connected building with multiple users:

---

[74] Edwards, 2016, p18 – 20 (Working Paper version at http://www.create.ac.uk/publications/privacy-security-and-data-protection-in-smart-cities-a-critical-eu-law-perspective/)
[75] Article 29 Working Party, 2014, p6; Rose et al, 2014, p26-29
[76] Edwards, 2016, p32

"…*it's quite straightforward if you've got something like a fitness band that connects to an app, and* you've got a single user who signs up to that app and has an account and it's in a kind of closed network, that is quite straightforward to put a privacy policy in and a consent process in place… *if you start thinking about connected buildings, where you've got everything from cameras to thermostats to infrared sensors light sensors, temperature sensors that are collecting data relating to lots and lots of different people,* it starts to get a lot more complicated in terms of getting consent from all those different people, and also in terms of those people knowing what data is being collected and what is being done with it." [Innes, Associate]

Creating effective consent mechanisms is not a job for lawyers or designers alone. At a higher level, there is an explicit turn the role of designers in regulation, as exemplified by privacy by design. Indeed, (PbD) is often cited as the solution to many challenges of IoT.[77] However, as we have argued elsewhere, to move PbD from theory into practice, a joint conceptual and practical approach is necessary.[78] We suggest turning to the user centric tools and approaches prevalent within the human computer interaction community.[79] Raising designer awareness of law is important, with new design tools being necessary to support this, like information privacy by design cards[80] or privacy design patterns.[81] We now turn to greater detail on the nature of PbD, both in theory and practice.

## Part III: Regulatory Solutions? The Role of Privacy by Design

Privacy by Design (PbD) as a policy tool has been discussed in EU and UK regulatory circles for some time.[82] State regulatory bodies like the UK Information Commissioner Office[83], the European Data Protection Supervisor (EDPS)[84], European Union Agency for Network and Information Security (ENISA)[85], and EU Article 29 Working Party all recognise the importance of PbD approaches.[86] The EDPS, for example, has stated that "*systems and software engineers need to understand and better apply the principles of privacy by design in new products and services across design phases and technologies*".[87] More specifically for the IoT, the Article 29 Working Party Opinion recommends "*Every stakeholder in the IoT should apply the principles of Privacy by Design and Privacy by Default*".[88]

---

[77] Danezis, G. et al *Privacy and Data Protection by Design – from policy to engineering* (Heraklion: European Network Information Security Agency, 2014); Brown, 2015

[78] Urquhart, L. and Rodden, T. 'A Legal Turn in Human Computer Interaction? Towards 'Regulation by Design' for the Internet of Things' (2016a) Forthcoming, *SSRN Working Paper* http://papers.ssrn.com/sol3/papers.cfm?abstract_id=2746467

[79] Urquhart L and Rodden 'New Directions in Information Technology Law: Learning from Human Computer Interaction' (2016b) *International Review of Law Computers and Technology* Forthcoming

[80] Luger, E. et al, 'Playing the Legal Card: Using Ideation Cards to Raise Data Protection Issues within the Design Process', (2015) *In Proceedings SIGCHI Conference Human Factors in Computer Systems (CHI'15)* (New York: ACM Press, 2015)*,* 457-466

[81] Colesky M., et al 'Critical Analysis of Privacy Design Strategies' (2016) In *2016 International Workshop on Privacy Engineering – IWPE'16*, (San Jose: IEEE 2016)

[82] Cavoukian, A. '7 Foundational Principles of Privacy by Design', (Ontario: Information and Privacy Commissioner of Ontario, 2011) Spiekermann, S. 'The Challenges of Privacy by Design' (2012) *Communications of the ACM (CACM)* 55 (7), 34-37.

[83] Information Commissioner Website, *Privacy By Design*, (2016) accessed at https://ico.org.uk/for-organisations/guide-to-data-protection/privacy-by-design/

[84] European Data Protection Supervisor, *Drones - Opinion of 26 November 2014*, (Brussels: EDPS, 2014); European Data Protection Supervisor, *eCall System - Opinion of 29 October 2013*, ((Brussels: EDPS, 2013)

[85] Danezis et al., 2014

[86] Article 29 Working Party, 2014

[87] European Data Protection Supervisor, *Towards a New Digital Ethics Opinion 4/2015*, (Brussels: EDPS, 2015) p10

[88] Article 29 Working Party Opinion, 2014, p21

The core idea is for designers of technology to consider privacy challenges as early as possible, ideally before a system is built or goes to market, in order to embed appropriate safeguards. In some regards, it aims to narrow the regulatory effectiveness gap created by slow legislative change and quick technological development. Article 25 (1) of the 2016 General Data Protection Regulation 2016[89] puts this concept into law and will be enforced across the EU from 25 May 2018.[90] Article 25 places obligations on data controllers to protect freedoms and rights of individuals, implement data protection principles and generally comply with requirements of the new law. This is done by employing safeguards during data processing (Article 4(2), GDPR, 2016).[91] Safeguards require adoption of appropriate technical and organisational measures that reflect the:

- *"state of the art,*
- *the cost of implementation*
- *the nature, scope, context and purposes of processing*
- *risks of varying likelihood and severity for rights and freedoms of natural persons posed by the processing"* (Article 25(1), GDPR, 2016)

Safeguards shall be considered at the *'time of the determination of the means for processing and at the time of the processing itself'* and the measures can consist of:

- "*minimising the processing of personal data,*
- *pseudonymising personal data as soon as possible,*
- *transparency with regard to the functions and processing of personal data,*
- *enabling the data subject to monitor the data processing,*
- *enabling the controller to create and improve security features.*" (Recital 78, GDPR, 2016)

In addition, by default, technical and organisational measures should be taken to ensure processing is:

- For *'personal data which are necessary for each specific purpose of the processing'*;
- Controlling the '*amount of personal data collected, the extent of their processing, the period of their storage and their accessibility*'
- That '*personal data are not made accessible without the individual's intervention to an indefinite number of natural persons.*' (Article 25(2), GDPR, 2016)

Compliance with Article 25 can be demonstrated by implementing organisational and technical measures, adopting internal policies, or accreditation through new certification mechanisms.[92] Legal compliance is important because a key attribute of the new GDPR is much larger fines. Data controllers or processors can now be charged up to the higher of €20m or 4% of global turnover (Article 83, GDPR, 2016) for failure to comply. We assert that drafting in the GDPR[93] of data controller and data processing are sufficiently broad to put those creating new

---

[89] Formerly Article 23 of draft
[90] Wood, S. 'GDPR Still Relevant for UK', (2016) UK Information Commissioner Office Blog, 7 July 2016 https://iconewsblog.wordpress.com/2016/07/07/gdpr-still-relevant-for-the-uk/
[91] Processing is very broad – see below.
[92] These are proposed but not yet developed.
[93] EU General Data Protection Regulation (GDPR), 2016, Recital 78 – (emphasis added) "When **developing, designing, selecting and using applications**, **services and products** that are based on the processing of personal data or process personal data to fulfil their task, **producers of the products, services and applications** should be encouraged to take into account the **right to data protection** when **developing and designing such products**, **services and applications** and, with due regard to the state of the art, to make sure that controllers and processors are able to fulfil their data protection obligations"

technologies at the forefront of compliance, including a range of system designers from manufacturers to third party services.[94]

A key stumbling block with PbD is how it might work in practice. A detailed understanding of the law and policy environment is not prevalent with engineers. Birnhack, Toch and Hadar (2014) have argued *"whereas for lawyers PbD seems an intuitive and sensible policy tool, for information systems developers and engineers it is anything but intuitive"* (p3). Similarly, Danezis et al ENISA Report on Privacy by Design Tools highlighted *"we observed that privacy and data protection features are, on the whole, ignored by traditional engineering approaches when implementing the desired functionality. <u>This ignorance is caused and supported by limitations of awareness and understanding of developers and data controllers as well as lacking tools to realise privacy by design.</u> While the research community is very active and growing, and constantly improving existing and contributing further building blocks, it is only loosely interlinked with practice"*.[95]

Legal commentators, like Jaap Koops and Leenes echo this, arguing guidance on PbD in practice is lacking[96], as does Brown who argues *"the specifics of implementation [for PbD] have so far only been developed to a limited extent"*.[97] Solutions are needed to bridge the gap between these two communities, as mentioned above. Jaap Koops and Leenes argue for focusing energy on communication between lawyers and designers, *"fostering the right mind-set of those responsible for developing and running data processing systems"*.[98] Indeed, law is not intuitive or accessible to non-lawyers, yet by calling for privacy by design, the law is mandating non-lawyers to be involved in regulatory practices. There is a need to engage, sensitise and guide designers on data protection issues on their own terms. We have made endeavours in this direction exploring how to practically do PbD, and support interaction between these the legal and design communities both conceptually[99] and practically using information privacy by design cards.[100]

Beyond this social level, complementary technical work addressing privacy challenges of systems presents a useful blueprint to doing PbD. Such approaches include privacy and security engineering,[101] usable privacy and security,[102] privacy enhancing technologies (PETS)[103], and most recently, Human Data Interaction.[104]

With usable privacy and security, for example, the goal is to create technical responses to regulatory challenges that are comprehensible and usable for end users. A broad church, usable privacy has work on increasing user control by setting machine readable permissions for data collection from devices/browsers (so called P3P)[105]; improving quality of notice and choice mechanisms by creating 'nutrition labels' for users to compare privacy policies[106], or nudging users towards more cautious sharing practices.[107] A combination of technical and social approaches to PbD are key.

---

[94] GDPR, 2016, Article 4(5) - "the natural or legal person, public authority, agency or any other body which alone or jointly with others determines the purposes and means of the processing of personal data"
[95] Danezis et al, 2014, p4
[96] Jaap Koops and Leenes, 2014, p161
[97] Brown, 2015, p26
[98] Jaap Koops and Leenes, 2014, p168
[99] Urquhart and Rodden, 2016a; Urquhart and Rodden, 2016b; Urquhart, L *Towards User Centric Regulation: Exploring the Interface Between IT Law and HCI*, (Nottingham: University of Nottingham/PhD Thesis, 2016)
[100] Luger et al, 2015
[101] Dennedy, M., Fox, J. and Finneran, T *Privacy Engineer's Manifesto*, (New York: Apress, 2014); Spiekermann, S., & Cranor, L.F. 'Engineering Privacy' (2009) *IEEE Transactions on Software Engineering* 35 (1), 67-82; Oliver, I. *Privacy Engineering*, (Independently Published, 2014); Anderson R., *Security Engineering*, (New York: Wiley, 2nd Ed, 2011),
[102] Iachello, G., and Hong, J. 'End User Privacy in Human Computer Interaction' (2007) *Foundations and Trends in Human Computer Interaction* 1(1), 1-137
[103] Camp, J., & Osorio, C., 'Privacy-Enhancing Technologies for Internet Commerce', *Trust in the Network Economy*: (Berlin: Springer-Verlag, 2003)
[104] Mortier, R. et al, 'Human-Data Interaction: The Human Face of the Data Driven Society',(2014) *SSRN Working Paper*, accessed at http://www.eecs.qmul.ac.uk/~hamed/papers/HDIssrn.pdf

We now briefly consider a few points around PbD in practice. Ideologically, the lawyers, view PbD positively because it exposes privacy risks earlier in the design process, allowing them to be addressed and avoiding 'back-pedalling down the line'. However, our experts argue PbD requires greater critique and reflection, including both communities sharing what works and what does not. At the abstract level, even the terms 'privacy' and 'by design' are disputed. For example, with privacy, some technologists question how to design for such a contested social value. As Iain frames it:

"*<u>My problem with privacy by design is I don't know what people mean by privacy anymore because all of the practices we carry out contradict the values</u>…you can't expect people to recover values, and when they do recover values, they go back to really old school values like Christian values or family values, or I don't think they know what privacy values are, I think they contradict them all the time, through the value propositions of which ubiquity in networked data offer them.*" [Iain, Professor]

Other technologists questioned what PbD actually means, like Kenneth who stated, "*I don't really think industry understand what headline PbD actually means, I don't think there has been a communication that has turned it into what does this mean?*". Instead, he argues it is better to understand how PbD will play out differently for specific application sectors and solutions, for example, with cars, cities and the home, as opposed to a generalised approach.

Some lawyers are concerned how PbD is framed as a solution. Whilst PbD may involve building in privacy enhancing tools like encryption into a system, more fundamentally it is important to reflect on if PbD is meant to support existing legal approaches to privacy, or to replace them with technical measures. The latter did not work well with copyright and digital rights management; hence Campbell warns:

"*The idea that you can substitute legal protections and balances with technological protections or workaround… I think that will be trickier…if it's [PbD] seen as part of a broader approach, as in one of the ways in which agreed legal standards can be implemented, so it doesn't add anything new in legal terms, it just takes, what I inelegantly call prevention instead of cure response, that has some potential*" [Campbell, Reader]

Another interesting finding is the parallels between privacy by design and security by design. Security is often not considered until later in the development cycle, retrospectively bolted on after the device been rushed to market as there are minimal motivations to thinking about security. In practice, this problem is particularly pronounced with SMEs and start-ups who are financially constrained. Technologist Hamish [Director and Managing Consultant, 30 years, Cybersecurity and Identity Management] captures this well:

---

[105] Cranor et al, 'The Platform for Privacy Preferences 1.0 (P3P)', (2002) *W3C Recommendation*
[106] Kelley, P.G et al, 'A Nutrition Label for Privacy', (2009) *Symposium on Usable Privacy and Security (SOUPS) Conference*, July 15-17 at https://cups.cs.cmu.edu/soups/2009/proceedings/a4-kelley.pdf
[107] Wang, Y. et al, 'A Field Trial of Privacy Nudges for Facebook' *In Proceedings SIGCHI Conference Human Factors in Computer Systems (CHI'14)* (New York: ACM Press, 2014)

*"…I think for start-ups and SMEs unless security is critical to their project, it largely gets overlooked because they are working on a limited budget, they have got to start producing revenue as soon as they can, and therefore it is a race to get your component to market as quickly as you possibly can, and it is very much seen, well there might be some security things, but we'll fix those in version 2…sadly, it never happens in version 2 and what does happen in v2.0 is a bit too late, because of some of the design decisions made in version 1"* [Hamish, Director and Managing Consultant]

Similarly, Ewan argues as start-ups lack the financial resources to obtain legal advice they are focusing on getting more investment to stay in business, not compliance. As he puts it *"the people that I speak to who are in [tech incubators], a lot of them all they are interested in really is raising money from investors, because that is what they need to live on, because they are living pretty much hand to mouth, so their interest is in winning business and getting investment…things like Privacy by Design and Privacy by Default, how they are anonymising data, what their data retention policies are, how they communicate with customers, that's just not on their radar…"* [Ewan, Partner]

Clearly, more critical reflection on PbD in practice is important. Understanding privacy as a value is an issue for designers, as it is a contested term and looking to users may not provide much clarity. Learning from similarities with security by design may be useful, as can questioning what PbD is doing - augmenting or replacing traditional legal approaches? For experts, the former is preferable, but importantly, like with IoT, PbD cannot be understood in the abstract, and focusing on how it manifests in specific sectors is important, instead of at a general level. Furthermore, commercial realities of limited financial and organisational resources, coupled with a different focus, especially for SMEs and start-ups, have to be factored into any workable notion of PbD.

## Part IV: Conclusions

Within this paper, we evaluated the regulatory challenges ambient domestic computing systems pose both conceptually and empirically. We provided insight into the longstanding visions of post-desktop computing, such as ubicomp and ambient intelligence and used these to conceptually situate the current trend of domestic IoT. We presented current framings of domestic IoT through analysis of practical experiences of leading experts in technology law and design. We considered conceptual legal challenges for IoT, and situated through observations of trends in the emerging IoT market. Practical regulatory challenges include designing effective consent mechanisms across heterogeneous devices, allowing users control over inferences from flows of information, and establishing parties with legal responsibilities in domestic IoT ecosystems. Equally, the regulatory solution of PbD whilst theoretically welcomed, faces practical implementation challenges. In practical terms, we observed issues such as lack of sector specific guidance and inadequate financial or organisational resources to enable businesses to do PbD in practice. Through analysis of earlier ambient domestic technologies we observe the importance of considering how a technology mediates a user's life in context and to respond accordingly. For PbD in the domestic context, this means creating design approaches that engage with values, like privacy, in the setting of the home. Neglecting the complex interactions and practices between users, services and devices risks moving ambient domestic computing a step closer to Bradbury's darker prophecies of the future home.

76. Wood, S. 'GDPR Still Relevant for UK', (2016) UK Information Commissioner Office Blog, 7 July 2016 https://iconewsblog.wordpress.com/2016/07/07/gdpr-still-relevant-for-the-uk/

77. Working Party on Communication Infrastructures and Services Policy, *Building Blocks for Smart Networks*, (Paris: OECD, 2013)

78. Wright, D. et al *Safeguards in a World of Ambient Intelligence*, (Netherlands: Springer, 2008)

Page 25 of 25